\documentclass[a4paper,11pt]{article}
\pdfoutput=1 

\usepackage{jheppub} 

\usepackage[T1]{fontenc} 
\usepackage[utf8]{inputenc}
\usepackage{cancel, multirow, amsmath}
\usepackage{graphicx}
\usepackage{caption}
\usepackage{subcaption}
\usepackage{float}
\usepackage{soul}
\usepackage{hyperref}
\hypersetup{
    linktocpage=true,
    colorlinks=true,
    linkcolor=blue,
    citecolor=green,
    pdftitle={Title},
    bookmarks=true}
\usepackage{cleveref}
\usepackage{xcolor}

\newcommand{\vinf}{V_{\text{inf}}}
\newcommand{\mpl}{M_{\text{pl}}}
\newcommand{\lb}{ \left( }
\newcommand{\rb}{ \right) }

\graphicspath{{figs/}}

\title{\boldmath TwInflation}

\author[a]{Kaustubh Deshpande,}
\author[b,c]{Soubhik Kumar}
\author[a]{and Raman Sundrum}
\affiliation[a]{Maryland Center for Fundamental Physics, Department of Physics, \\University of Maryland, College park, MD 20742, USA}
\affiliation[b]{Berkeley Center for Theoretical Physics, Department of Physics, \\University of California, Berkeley, CA 94720, USA}
\affiliation[c]{Theoretical Physics Group, Lawrence Berkeley National Laboratory, Berkeley, CA 94720, USA}

\emailAdd{ksd@umd.edu}
\emailAdd{soubhik@berkeley.edu}
\emailAdd{raman@umd.edu}

\abstract{
The general structure of Hybrid Inflation remains a very well-motivated mechanism for lower-scale cosmic inflation in the face of improving constraints on 
the tensor-to-scalar ratio.
However, as originally modeled, the ``waterfall'' field in this mechanism gives rise to a hierarchy problem ($\eta-$problem) for the inflaton after demanding standard effective field theory (EFT) control. We modify the hybrid mechanism and incorporate a discrete ``twin'' symmetry, thereby yielding 
 a viable, natural and EFT-controlled model of non-supersymmetric low-scale inflation, ``Twinflation''.
Analogously to Twin Higgs models, the 
 discrete exchange-symmetry with a ``twin'' sector
  reduces quadratic sensitivity in the inflationary potential to ultra-violet physics, at the root of the hierarchy problem.
The observed phase of inflation takes place on a hilltop-like potential but without  fine-tuning of the initial inflaton position in field-space. We also show that all parameters of the model can take natural values, below any associated EFT-cutoff mass scales and field values,
thus ensuring straightforward theoretical control. We discuss the basic phenomenological considerations and constraints, as well as possible future directions.

}

\begin{document} 

\preprint{UMD-PP-021-01}

\maketitle
\flushbottom

\section{Introduction}
\label{sec:intro}

Cosmic inflation (see~\cite{Baumann:2009ds} for a review) is an attractive and robust framework for helping to explain the state of the early universe, resolving issues such as the horizon problem, the flatness problem, and the origin of primordial fluctuations. It can be implemented minimally by the slow rolling of a single real scalar field, the inflaton $(\phi)$, along its nearly flat potential ($V(\phi)$). But, this requires the inflaton to be significantly lighter than the Hubble scale, which gives rise to a hierarchy problem known as the ``$\eta-$problem'' (see e.g.~\cite{Baumann:2014nda}). 

Furthermore, the observations so far \cite{Planck2018Inflation} seem to rule out or strongly constrain some of the simplest forms of $V(\phi)$, originating from straightforward and natural microscopic models explaining the lightness of the inflaton. They typically predict  a large tensor-to-scalar ratio, $r\gtrsim 0.01$, and hence a high scale of inflation. But, with the non-observation of primordial tensor fluctuations to date, the data seems to hint towards lower-scale inflation. The upcoming and near-future proposed experiments like BICEP Array \cite{BICEP_Array_Hui:2018cvg}, Simons Observatory \cite{Simons_Observatory_Ade:2018sbj}, CMB-S4 \cite{CMB_S4_Abazajian:2019eic}, LiteBIRD \cite{LiteBIRD_Hazumi:2019lys}, and PICO \cite{PICO_Hanany:2019lle}, will be able to measure $r \gtrsim 10^{-3}$, corresponding to $H \gtrsim 5\times 10^{12}$ GeV.
It is therefore interesting to reconsider the structure of inflationary dynamics, especially keeping the $\eta-$problem in mind, to see whether observable $r$ is a robust prediction or whether extremely small $r$ can be readily achieved.

Indeed, inflation may well take place at a much lower scale than above, i.e. with $H \ll 10^{12}$ GeV, with unobservably small tensor fluctuation at these near-future experiments, although, realizing such low-scale inflation with a simple single-field model is typically fine-tuned. 
This fine-tuning can come in the form of the potential, the model parameters, and also the initial conditions (see e.g.~\cite{Goldwirth:1991rj, Dine:2011ws, Brandenberger:2016uzh,Linde:2017pwt,Chowdhury:2019otk}).
On the other hand, multi-field inflation, i.e. with the field(s) orthogonal to inflaton playing an important dynamical role in (ending) inflation, can help in the model building for low-scale inflation. The classic example of this is Hybrid Inflation \cite{Linde:1993cn}. Here, the inflaton couples to a ``waterfall'' field ($\sigma$) in such a way that $\sigma$ has a $\phi$-dependent mass term. During inflation, the much heavier $\sigma$ is fixed at $0$, while $\phi$ performs the slow roll. As the inflaton rolls past a critical field value, $\sigma$ becomes tachyonic and rapidly rolls down to the global minimum of the potential. This fast rolling along the ``waterfall'' on the inflationary trajectory ends inflation by releasing the vacuum energy in the $\sigma$ field. Hybrid inflation exhibits a separation of roles with the space-time expansion during inflation dominantly driven by vacuum energy in $\sigma$, and the slow-roll ``clock'' provided by $\phi$, which helps in realizing low-scale inflation as we will review in Sec.~\ref{sec:hybrid_inflation_review}. This provides a mechanism generating an effective inflationary trajectory with an abrupt drop in vacuum energy, which is difficult to realize from a single-field perspective.
However, as we will review in Sec.~\ref{sec:hybrid_inflation_review}, hybrid inflation needs fine-tuning in the model parameters to achieve radiative stability and EFT control. We will address this issue in the present work and build an EFT-controlled and natural low-scale inflationary model.

The primary challenge offered by the hybrid inflation paradigm towards building a microscopic model is the following: $\phi$ needs to be a light real scalar, but with sufficiently strong non-derivative coupling with the heavy $\sigma$ field as required for the waterfall effect. Even if $\phi$ is modeled as a pseudo-Nambu Goldstone boson (pNGB) of a global symmetry, its coupling with $\sigma$ explicitly breaks the symmetry and induces quadratic sensitivity in the effective inflationary potential to the ultra-violet (UV) physics. Hence, we need some extra ingredient to achieve naturalness in hybrid inflation. This issue is similar to the case of the light Higgs boson as required in the Standard Model (SM) in the presence of its Yukawa and gauge couplings. This, hence, motivates one to apply different particle physics mechanisms explored in the literature to address the hierarchy problem of the SM Higgs boson, to the case of hybrid inflation mentioned above.
There are various supersymmetric constructions of hybrid inflation, see e.g. \cite{Copeland:1994vg, Dvali:1994ms, Binetruy:1996xj, Halyo:1996pp, Kallosh:2003ux}.
Little Inflaton \cite{Kaplan:2003aj, ArkaniHamed:2003mz} is also one such proposal addressing the issue of naturalness in hybrid inflation based on the Little Higgs mechanism \cite{Little_Higgs}. This makes use of ``collective symmetry breaking'' to protect the inflaton potential from the radiative contributions sourced by its coupling with the waterfall field.
See also \cite{Sundrum:2009ii, Ross:2016hyb, Kaloper:2020jso, Carta:2020oci} for more proposals aimed at building such a radiatively stable, EFT-controlled and viable model for hybrid inflation.

Twin Higgs \cite{Chacko:2005pe} is another mechanism proposed to address the (little) hierarchy problem of the SM Higgs boson. 
Here, the light scalar is protected from radiative corrections sourced by its non-derivative couplings by using a discrete  symmetry, with a symmetry-based cancellation of 1-loop quadratic divergences. Inspired by this, in the present work, we make use of a $\mathbb{Z}_2$-symmetry structure to build a quite simple, natural and EFT-controlled model of hybrid inflation, which we will call  ``Twinflation''.\footnote{We thank N. Craig, S. Koren and T. Trott for giving us permission to re-use this name, first used by them in the different setting of Ref.~\cite{Craig:2016lyx}.} As we will see in Sec.~\ref{sec:the_model}, Twinflation can naturally give rise to a viable model of inflation, with a red tilt in the primordial scalar fluctuations consistent with the observations \cite{Planck2018Inflation}, and with the inflationary Hubble scale as low as $\sim 10^7$ GeV.

Low-scale inflation and the consequent reheating, apart from explaining the smallness of yet-unobserved primordial tensor fluctuations, can also be motivated from other particle physics considerations. For example, if QCD axions or axion-like particles constitute (a significant fraction of) cold dark matter (CDM) and if Peccei-Quinn (PQ) symmetry is broken during inflation, low-scale inflation is favored to avoid CDM isocurvature constraints (see e.g.~\cite{Axion_Cosmology_Review_Marsh:2015xka,ALPs_isocurvature_Diez-Tejedor:2017ivd,Planck2018Inflation}). Such inflationary scenarios are also often invoked so that heavy, unwanted relics e.g. monopoles, moduli, gravitino, which might be generated by the UV physics (see e.g.~\cite{GravitinoProblem_Ellis:1982yb, GravitinoProblem_Ellis:1984eq,  GravitinoProblem_Murayama_etal, ModuliProblem_Randall:1994fr})
are diluted away/not reheated.\footnote{We note that it is also possible to avoid reheating heavy relics just by requiring a low reheating temperature while still having a high-scale inflation.} Furthermore, for sufficiently low inflationary scales, we can have complementary terrestrial particle physics probes of inflation and reheating, such as at current and future collider experiments, see e.g. \cite{Bezrukov:2009yw, Allahverdi:2010zp, Boehm:2012rh, Bramante:2016yju}.

The paper is organized as follows. In Sec.~\ref{sec:hybrid_inflation_review}, we review the basic mechanism of hybrid inflation, also reviewing that it requires fine-tuning of parameters to achieve radiative stability and EFT control, the criteria of which we also explain. In Sec.~\ref{sec:soft_coupling}, we present a simple variant of hybrid inflation with a {\it soft} (dimensionful) waterfall coupling, and show that even this suffers from a similar naturalness problem as before. In Sec.~\ref{sec:effective_single_field_inflation}, we describe the effective single-field inflation with the massive waterfall field integrated out. Here, we also introduce a simplifying notation for the effective inflationary potential that arises quite generically from hybrid inflation (irrespective of its naturalness) using which we can estimate the inflationary observables and constrain some model parameters. In Sec.~\ref{sec:the_model}, we construct the Twinflation model, starting with a simple renormalizable version, analysing its radiative stability and EFT consistency, and then presenting a more complete version realizing the pNGB structure of the inflaton. In Sec.~\ref{sec:domain_wall_problem}, we discuss a simple way to address the cosmological domain wall problem related to the spontaneous breaking of a (simplifying but non-essential) $\sigma$-parity at the end of inflation, via a small explicit breaking. We conclude in Sec.~\ref{sec:discussion}.

\section{Hybrid inflation and naturalness}
\label{sec:hybrid_inflation_review}

The basic mechanism of hybrid inflation can be described by the following simple variant~\cite{Lyth:1996kt} of the original potential in~\cite{Linde:1993cn}:
\begin{equation}
\label{eq:basic_hybrid_inflation_on_hilltop}
	V(\phi, \sigma) = \vinf + v(\phi) + \frac{1}{2}M_\sigma^2 \sigma^2 + \frac{1}{4}\lambda_\sigma \sigma^4 - \frac{1}{2}  g \phi^2 \sigma^2 + \dots .  
\end{equation}
Here, $\phi$ is the slowly rolling inflaton and $\sigma$ is the ``waterfall'' field whose dynamics ends inflation.
Inflation starts at small $\phi$, with $0<g \phi^2 < M_\sigma^2$, such that the minimum in the $\sigma$ direction is at $\sigma = 0$.
The ellipsis in Eq.~\eqref{eq:basic_hybrid_inflation_on_hilltop} includes higher-dimensional interaction terms ensuring global stability of the potential at large field values.  
A crucial ingredient of the hybrid inflation mechanism is that during inflation the $\sigma$-mass is bigger than both the $\phi$-mass and the Hubble scale.
This ensures that $\sigma$ remains localized at $\sigma=0$, and does not play any role until the end of inflation. Therefore, during inflation, i.e. for $g \phi^2 < M_\sigma^2$, $V(\phi, \sigma)$ in Eq.~\eqref{eq:basic_hybrid_inflation_on_hilltop} effectively reduces to
\begin{align}
\label{eq:effective_inf_potential_Linde}
	V_{\rm eff}(\phi)\approx \vinf + v(\phi).
\end{align}
For $|v(\phi)|\ll \vinf$, this implies that the detailed dynamics of the inflaton is governed by $v(\phi)$, while the vacuum energy $\vinf$ dominantly drives the spacetime expansion. We will see that the relaxation of $V_{\rm inf}$  to zero, as needed at the end of inflation, can be triggered by $\sigma$ dynamics, rather than purely the single-field rolling of $\phi$. 
The crucial separation of roles between $v$ and $V_{\rm inf}$ is one of the primary reasons why the waterfall mechanism allows for consistent low-scale models of inflation. 

As inflation progresses, $\phi$ slowly rolls down its potential $v(\phi)$, i.e. towards larger $\phi$. As it crosses a critical value $\phi_*= \frac{M_\sigma}{\sqrt{g}}$ 
(assumed to be smaller than the minimum of $v(\phi)$), the effective mass-squared for $\sigma$ switches sign. Consequently, the now-tachyonic  $\sigma$ rapidly rolls down to its new minimum. This \emph{fast} rolling of the waterfall field violates the \emph{slow-}roll conditions and ends inflation by releasing the inflationary vacuum energy, $\vinf$. 
The two fields finally settle into the global minimum which can be characterized by some $ \phi_{\rm min}$ with $\sigma_{\rm min} = \sqrt{\frac{g \phi_{\rm min}^2 - M_\sigma^2}{\lambda_\sigma}}$.  
Demanding a negligible vacuum energy in the post-inflationary era fixes
\begin{align}
\label{eq:Vinf_waterfall}
	\vinf = 3 H^2 \mpl^2 \approx \frac{\lb g \phi_{\rm min}^2 - M_\sigma^2 \rb^2}{4 \lambda_\sigma} = \frac{\lb 1 - \phi_{\rm min}^2/\phi_*^2 \rb^2}{4} \frac{M_\sigma^4}{\lambda_\sigma} \sim  \mathcal{O}(1) \frac{M_\sigma^4}{\lambda_\sigma}.
\end{align}
In the last step above, we have considered that the ellipsis in Eq.~\eqref{eq:basic_hybrid_inflation_on_hilltop} fixes the global minimum in $\phi$ only $\mathcal{O}(1)$ away from $\phi_*$, i.e. $\phi_* \sim \mathcal{O}(\phi_{\rm min}) $. 
This is also so that there is no tuning required in the initial inflaton field location (see also Sec.~\ref{sec:effective_single_field_inflation}). 
As we will see in Sec.~\ref{subsec:U(1)_UV_completion}, all these aspects can be easily realized with $\phi$ being a pNGB of a global symmetry and consequently its couplings taking trigonometric forms. 

In the original hybrid inflation model \cite{Linde:1993cn}, $v(\phi) = + \frac{1}{2} m_\phi^2 \phi^2$ along with an opposite choice of signs in the potential in Eq.~\eqref{eq:basic_hybrid_inflation_on_hilltop} for the $M_\sigma^2$ and $g$ terms, allowing inflation to start at large $\phi$. 
This convex form of $v(\phi)$ in hybrid inflation, however, leads to blue tilt in the power spectrum of the primordial scalar perturbations (after respecting the constaint on tensor-to-scalar ratio) which is strongly disfavored by the Planck data \cite{Planck2018Inflation}. In order to get the observed red tilted spectrum, we will consider a hilltop-like $v(\phi)$ \cite{Lyth:1996kt} with inflation happening somewhat near its maximum. In Sec.~\ref{sec:effective_single_field_inflation}, we will see that no tuning is required in the initial inflaton field value to achieve this.
A simple example of such a potential is
\begin{equation}
	\label{eq:hilltop_v(phi)}
	v(\phi) = - \frac{1}{2} m_\phi^2 \phi^2 + \frac{\lambda_\phi}{4} \phi^4 + \dots ,
\end{equation}
which has a hilltop at $\phi = 0$. The ellipsis above refers to sub-dominant higher-dimensional terms in $\phi$.

\subsection{Naturalness considerations}

In high-scale models of inflation, the inflaton field typically traverses super-Planckian field distances \cite{LythBound}, requiring special UV structures to ensure the consistency of the inflationary effective field theory,  e.g. as in \cite{Biaxion_KNP}. Here, for our lower-scale inflation, we will aim to have a more straightforward EFT consistency. In particular,
we will be aiming to construct a low-scale model of hybrid inflation where
\begin{itemize}
\item all the parameters take natural (or bigger) values,
\item all the relevant mass scales and field values are smaller than the respective EFT cutoff(s),
\item the EFT cutoff(s) is (are) sub-Planckian.
\end{itemize}
In the following, we will examine the naturalness of hybrid inflation, in light of the above requirements, first for the original model in Eq.~\eqref{eq:basic_hybrid_inflation_on_hilltop} (with a hilltop structure of $v(\phi)$) and then in Sec.~\ref{sec:soft_coupling} for our 
simple modification with a {\it soft} waterfall coupling.

The non-derivative coupling with the waterfall field in Eq.~\eqref{eq:basic_hybrid_inflation_on_hilltop} badly breaks shift symmetry of the inflaton and radiatively generates quadratic sensitivity in $m_\phi^2$ to the UV cutoff scale\footnote{More precisely, $\Lambda$ should be thought of as a placeholder for the mass of some heavy field.} $\Lambda$:
\begin{equation}
	\left(\delta m_\phi^2\right)_{\text{1-loop}} \sim \frac{g \Lambda^2}{16 \pi^2}.
\end{equation}
In order to satisfy naturalness in $m_\phi^2$, we require 
\begin{equation}
\label{eq:mphi_sq_1loop_quad_div}
	\left(\delta m_\phi^2\right)_{\text{1-loop}} \lesssim \left(m_\phi^2\right)_{\rm tree}  ~~ \textrm{i.e.} ~~  \Lambda^2 \lesssim \lb 16 \pi^2 \eta \rb \frac{H^2}{g},
\end{equation} 
implying that the UV cutoff $\Lambda$ cannot be arbitrarily large.
Here $\eta \equiv \mpl^2 \frac{\partial^2_\phi V(\phi,\sigma)}{V(\phi,\sigma)} \ll 1$ is the slow-roll parameter during inflation, with $(m_\phi^2)_{\rm tree} \sim \eta H^2$. Furthermore, the requirement that $\sigma$ is not dynamical during inflation, i.e. it being frozen at $\sigma = 0$, implies its effective mass should be bigger than the Hubble scale, 
\begin{equation}
\label{eq:msig_basic_hybrid_inf}
	M_{\sigma,\rm eff}^2 \equiv M_\sigma^2 - g \phi_0^2 \sim \mathcal{O}(1) \cdot g \phi_0^2 \gtrsim H^2,
\end{equation}
where $\phi_0$ denotes a typical inflaton field value during inflation and $M_{\sigma,\rm eff}^2 \sim M_\sigma^2 \sim \mathcal{O}(1) \cdot g \phi_0^2$. To satisfy conditions in  Eq.~\eqref{eq:mphi_sq_1loop_quad_div} and \eqref{eq:msig_basic_hybrid_inf},
we need 
\begin{equation}
	\phi_0^2 \gtrsim \frac{\Lambda^2}{16\pi^2\eta}.
\end{equation}
Since the observed tilt of the primordial perturbations gives $\eta\sim 10^{-2}$,
this demands inflaton field displacement bigger than the UV scale, i.e.
\begin{equation}
\label{eq:inflaton_displacement_bigger_than_cutoff}
	\phi_0 \gtrsim \Lambda.
\end{equation}
However, this is only marginally consistent with our requirements above, and we cannot take $\phi_0\ll \Lambda$ as desired.

Furthermore, even marginally satisfying validity of the EFT, i.e. $\phi_0 \sim \Lambda$ in Eq.~\eqref{eq:inflaton_displacement_bigger_than_cutoff}, we need to satisfy $M_{\sigma, \textrm{eff}}^2 \sim H^2$ in Eq.~\eqref{eq:msig_basic_hybrid_inf}. 
However, using Eq.~\eqref{eq:Vinf_waterfall}, this then requires the post-inflationary $\sigma$-VEV to be $\sim \mpl$:
\begin{equation}
\label{eq:sigma_VEV}
\langle \sigma \rangle_{\rm post-inf.}^2 \sim \frac{M_\sigma^2}{\lambda_\sigma} \sim \mpl^2 \frac{H^2}{M^2_\sigma} \sim \mpl^2 ,
\end{equation}
which is against our EFT requirements of sub-Planckian field values mentioned earlier.
In detail, $\langle\sigma^2 \rangle_{\rm post-inf.} = \frac{g \phi_{\rm min}^2 - M_\sigma^2}{\lambda_\sigma} = \frac{M_\sigma^2}{\lambda_\sigma} \lb \frac{\phi_{\rm min}^2}{\phi_*^2} -1 \rb$, and hence $\langle\sigma^2 \rangle_{\rm post-inf.} < \frac{M_\sigma^2}{\lambda_\sigma}$ is possible implying a slightly sub-Planckian $\sigma$-VEV. However, this is only marginal, and we would have a greater confidence in the EFT-control if the $\sigma$-VEV is \emph{parametrically} lower than $\mpl$. 

Thus, the only way to construct a consistent hybrid inflation model with Eq.~\eqref{eq:basic_hybrid_inflation_on_hilltop}, which is under EFT control, is with fine-tuning in $m_\phi^2$, i.e. with fine cancellations between $m^2_{\phi, \rm tree}$ and $\delta m^2_{\phi, \rm 1-loop}$. Only at the cost of such a tuning, can we satisfy $\phi_0 < \Lambda$.

\subsection{Allowing for different cutoff scales}

Since the quadratic sensitivity of $m_\phi^2$ at 1-loop comes due to the $\sigma$ field running in the loop, another solution one may try is allowing for different cutoff scales for $\phi$ and $\sigma$, i.e. $\Lambda_\phi$ and $\Lambda_\sigma$, respectively. 
This can come about if $\phi$ and $\sigma$ belong to two different sectors with different physical
scales involved in their UV completions. A familiar but dramatic example is given by the chiral Lagrangian description of composite pions of QCD, cut off by the GeV hadronic scale, 
while light leptons and gauge fields interacting
with these pions have a much higher cutoff. 

With a choice
\begin{equation}
	\label{eq:different_cutoffs}
	\Lambda_\phi \gtrsim \phi_0 \gtrsim \Lambda_\sigma,
\end{equation}
one may evade Eq.~\eqref{eq:inflaton_displacement_bigger_than_cutoff} while still ensuring EFT control in the $\phi-$sector. Now, we examine if hybrid inflation 
satisfies naturalness for all couplings, all scales being sub-Planckian and also smaller than the respective cutoffs, i.e. $m_\phi, \phi_0 \lesssim \Lambda_\phi$ and $M_{\sigma}, \langle \sigma \rangle  \lesssim \Lambda_\sigma$.
The radiative corrections to $m_\phi^2$ now are
\begin{equation}
	\lb \delta m_\phi^2 \rb_{\rm 1-loop} \sim \frac{g \Lambda_\sigma^2}{16 \pi^2} \gtrsim \frac{g \langle \sigma \rangle^2}{16 \pi^2} \sim \frac{H^2 \mpl^2}{16 \pi^2 \phi_0^2},
\end{equation}
where we use $\Lambda_\sigma \gtrsim \langle \sigma \rangle$ and $\langle \sigma \rangle \sim \frac{H \mpl}{\sqrt{g} \phi_0}$ following Eq.~\eqref{eq:sigma_VEV}. 
Now, we can see that 1-loop naturalness in $m_\phi^2$, i.e. $\lb \delta m_\phi^2 \rb_{\rm 1-loop} \lesssim m_\phi^2 \sim \eta H^2$, can only be satisfied with
\begin{equation}
	\label{eq:phi0_gtrsim_Mpl}
	\phi_0 \gtrsim \mpl,
\end{equation}  
which is against our requirements to realize a truly low-scale hybrid inflation model.

Thus, even allowing for separate cutoffs, hybrid inflation is still not naturally in EFT control.

\section{Hybrid inflation with a soft ``waterfall'' coupling}
\label{sec:soft_coupling}

The naturalness problem described in Sec.~\ref{sec:hybrid_inflation_review}  stems from the quadratic UV scale sensitivity in $m_\phi^2$. One of the simplest solutions is to have only a soft shift symmetry breaking for $\phi$, i.e. a dimensionful $\phi-\sigma$ interaction, e.g.
\begin{equation}
\label{eq:soft_coupling_hybrid_inflation}
V(\phi, \sigma) = \vinf + \lb - \frac{m_\phi^2}{2} \phi^2 + \frac{\lambda_\phi}{4} \phi^4 + \dots \rb +
\lb \frac{M_\sigma^2}{2} \sigma^2 + \frac{\lambda_\sigma}{4} \sigma^4 \rb - \frac{ \mu \phi}{2}  \sigma^2 + \dots.
\end{equation}
Here, during inflation, i.e. for $\mu \phi < M_\sigma^2$, $\sigma$ remains localized at $\sigma = 0$, thus giving the same effective inflationary potential as Eq.~\eqref{eq:effective_inf_potential_Linde}. The ellipsis after the last term in Eq.~\eqref{eq:soft_coupling_hybrid_inflation} above, as in Eq.~\eqref{eq:basic_hybrid_inflation_on_hilltop}, includes higher-dimensional interaction terms which ensure that the global minimum in $\phi$ is only $\mathcal{O}(1)$ away from the critical value $\phi_* = \frac{M_\sigma^2}{\mu}$. As $\phi$ rolls down past $\phi_*$, the waterfall in $\sigma$ is triggered, thus ending inflation by releasing the inflationary vacuum energy $\vinf \sim \mathcal{O}(1) \frac{M_\sigma^4}{\lambda_\sigma}$, similarly to Eq.~\eqref{eq:Vinf_waterfall}. As mentioned before, this parametric form of $\vinf$ along with $\phi_{\rm min} \sim \mathcal{O}( \phi_*)$ can be explicitly realized in the 
pNGB realization of the inflaton which we detail in 
 Sec.~\ref{subsec:U(1)_UV_completion}.

\subsection{Naturalness considerations} 

The soft coupling $\mu$ generates only a logarithmic cutoff sensitivity in $m_\phi^2$: 
\begin{equation}
	(\delta m_\phi^2)_{\rm 1-loop}\sim \frac{\mu^2 \ln\Lambda}{16\pi^2}.
\end{equation}
As in the previous case, demanding that the loop-induced inflaton mass is smaller than its tree-level mass, i.e. $\frac{\mu^2}{16 \pi^2} \lesssim \eta H^2$ (taking $\ln\Lambda \sim \mathcal{O}(1)$), and that $\sigma$ is non-dynamical during inflation, i.e. $M_{\sigma,\textrm{eff}}^2 \sim \mu \phi_0 \gtrsim H^2$,  we get
\begin{align}\label{mu}
	\frac{H}{\phi_0} \lesssim \frac{\mu}{H} \lesssim 4\pi \sqrt{\eta} \sim \mathcal{O}(1).
\end{align}
Therefore, at the first sight, there is no constraint such as $\phi_0 \gtrsim \Lambda$ as before. 
However, the $\mu$ term in Eq.~\eqref{eq:soft_coupling_hybrid_inflation} 
also generates a quadratically divergent $\phi$-tadpole: 
\begin{align}
	\label{eq:quadratically_divergent_tadpole}
	V(\phi,\sigma) \ni \frac{\mu \Lambda^2}{16 \pi^2} \phi.
\end{align}
Indeed, the soft waterfall coupling breaks $\phi \rightarrow -\phi$ symmetry allowing for a tadpole like above. Although it is possible for the theory to have a larger tadpole, e.g. $\Lambda^3 \phi$, but it is \emph{natural} for it to have the above radiatively generated value.
We take $\mu \ll \Lambda$ to characterize the small breaking of $\phi \rightarrow -\phi$ symmetry in any coupling of the model.
The tadpole in Eq.~\eqref{eq:quadratically_divergent_tadpole} can be absorbed in Eq.~\eqref{eq:soft_coupling_hybrid_inflation} 
with a large shift in the $\phi$ field:
\begin{equation}
	\delta \phi \sim \frac{\mu \Lambda^2}{16 \pi^2 m_\phi^2} \sim \frac{\mu \Lambda^2}{16 \pi^2 \eta H^2} \sim \frac{\mu \Lambda^2}{H^2}.
\end{equation}
Such a large shift in $\phi$, however, also gives large contributions to other terms in Eq.~\eqref{eq:soft_coupling_hybrid_inflation}, e.g.
\begin{equation}
	\frac{\delta M_\sigma^2}{M_{\sigma, \textrm{eff}}^2} \sim \frac{\delta \phi}{\phi_0} \sim \frac{\mu \Lambda^2}{H^2 \phi_0} \sim \frac{M_{\sigma, \textrm{eff}}^2}{H^2} \frac{\Lambda^2}{\phi_0^2}.
\end{equation}
We can see from above that, in order for naturalness in $M_\sigma^2$ (and also to allow for waterfall transition), i.e. for $\delta M_\sigma^2 \lesssim M_{\sigma, \textrm{eff}}^2$, we need
\begin{equation}
	\frac{\phi_0^2}{\Lambda^2} \gtrsim \frac{M_{\sigma, \textrm{eff}}^2}{H^2} \gtrsim 1.
\end{equation} 
This again implies $\phi_0 \gtrsim \Lambda$, which is in contradiction with the EFT requirements stated earlier.

\subsection{Allowing for different cutoff scales}

Allowing even for different cutoff scales in this hybrid inflation model with soft coupling, 
we get a similar result as Eq.~\eqref{eq:phi0_gtrsim_Mpl}. 
The radiative corrections to $M_\sigma^2$ here are
\begin{equation}
	\lb \delta M_\sigma^2 \rb_{\rm 1-loop} \sim \frac{\lambda_\sigma \Lambda_\sigma^2}{16 \pi^2} + \frac{\mu^2 \Lambda_\sigma^2}{16 \pi^2 m_\phi^2}.
\end{equation}
Naturalness for the first term on the right hand side above, as before, demands 
$ \langle \sigma \rangle \lesssim \Lambda_\sigma \lesssim 4 \pi \langle \sigma \rangle$, now with $\langle \sigma \rangle \sim \frac{H \mpl}{\sqrt{\mu \phi_0}}$. In order to satisfy naturalness for the second term (sourced by quadratically divergent $\phi$-tadpole), i.e.
\begin{equation}
	1 \gtrsim \frac{\mu^2 \Lambda_\sigma^2}{16 \pi^2 m_\phi^2 M_{\sigma}^2} \gtrsim 
	\frac{\mu \langle \sigma \rangle^2}{H^2 \phi_0} \sim  \frac{\mpl^2}{\phi_0^2},
\end{equation}
we again need
\begin{equation}
	\phi_0 \gtrsim \mpl.
\end{equation}

Thus, we see that with either marginal or soft $\phi-\sigma$ coupling, even with different cutoffs for the inflaton and the waterfall field, if we demand EFT control (i.e. all scales being smaller than the respective cutoffs) and sub-Planckian physics, the only way to have a consistent hybrid inflation model is with fine-tuning of the relevant parameters, $m_\phi^2$ or $M_\sigma^2$ as discussed in this and the previous section.
This suggests that in order to build a natural model for hybrid inflation, we need some significant new mechanism to entirely get rid of the quadratic UV-sensitivity in the inflaton potential coming from its necessarily non-derivative coupling to the waterfall field.


\section{Effective single-field inflation}
\label{sec:effective_single_field_inflation}

The models described in Sec.~\ref{sec:hybrid_inflation_review} and \ref{sec:soft_coupling} cannot give rise to consistent hybrid inflation under EFT control without fine-tuning of parameters. Before we propose such a natural model for hybrid inflation in Sec.~\ref{sec:the_model}, in this section we first focus on  effective single-field inflation with the massive waterfall field integrated out. We also introduce here a simplifying notation for the effective inflationary potential that arises quite generically from hybrid inflation. As we will see, 
this simplified single-field analysis allows us to
easily estimate the inflationary observables and use them to constrain the effective model parameters, even without knowing the detailed form of the full potential.
This ``satellite view'' will be helpful later in Sec.~\ref{sec:the_model} by simply identifying the realistic parts of parameter space deserving a fuller analysis.

The waterfall field, although with a $\phi$-dependent mass, still remains heavier than $H$ throughout inflation, except at the end of inflation when $M_\sigma^2(\phi)$ passes through zero. Thus, prior to the end of inflation we can integrate it out and get an effective single-field description in terms of $\phi$.
%
Hybrid inflation quite generically gives this effective single-field inflationary potential 
in the form of Eq.~\eqref{eq:effective_inf_potential_Linde}, which varies as some function $v(\phi)$ with a large vacuum energy offset $\vinf$. In this section, we introduce a simplifying notation with 
\begin{equation}
	\label{eq:v(phi)_simple_form}
	v(\phi) = V_0 \cdot F\left(\frac{\phi}{f}\right),
\end{equation}
where $V_0$ controls the magnitude, while the shape is specified by a dimensionless function $F$. The effective inflationary potential then has the following form:
\begin{equation}
	\label{eq:effective_single_field_inflation}
	V_{\rm eff}(\phi) = \vinf + V_0 \cdot  F\left(\frac{\phi}{f}\right) ~~ ; ~~ \vinf \gg V_0.
\end{equation}
The hilltop-like $v(\phi)$ that we considered earlier in Eq.~\eqref{eq:hilltop_v(phi)} has the form as in Eq.~\eqref{eq:v(phi)_simple_form}. We will also show later how this simple form arises generically from a more complete hybrid inflation model in Sec.~\ref{sec:the_model} where the inflaton is realized as a pNGB, and where $F\left(\frac{\phi}{f}\right)$ takes a trigonometric form.

The main benefit of using this simplifying notation is that, 
assuming the function $F$ and its derivatives are $\sim \mathcal{O}(1)$ during inflation,
which is also the case in the model that we discuss later in Sec.~\ref{sec:the_model},
we can obtain general expressions for inflationary observables as shown below, even without specifying the explicit form of $F$. We assume that inflation starts\footnote{More precisely, when the largest scales observable today exit the horizon during inflation.} at $\phi_i$ which is somewhat near the hilltop of $F\left(\frac{\phi}{f}\right)$ as preferred by the data \cite{Planck2018Inflation}, and ends at $\phi_e$ by a waterfall transition along the $\sigma$ field. 
Then, the slow-roll inflation parameters are\footnote{The slow roll parameters $\epsilon, \eta$ as defined above are, in general, functions of $\phi$. However, unless an explicit functional argument is shown, they refer to the parameters evaluated at an epoch when the largest scales observable today exit the horizon during inflation, normally $\sim$50-60 e-folds before the end of inflation.} 
\begin{equation}
\label{eq:slow_roll_inflation_param}
\begin{split}
&\eta \equiv  \frac{V^{\prime \prime}}{V} \mpl^2 \sim \frac{V_0}{\vinf} \frac{\mpl^2}{f^2} \ , \ 
\epsilon \equiv \frac{1}{2}\left(\frac{V^\prime}{V}\right)^2 \mpl^2  \sim \eta^2 \frac{f^2}{\mpl^2} , \\
&A_s \equiv \frac{1}{8 \pi^2}\frac{H^2}{\mpl^2} \frac{1}{\epsilon} \sim \frac{10^{-2}}{\eta^2} \frac{H^2}{f^2}  \ , \ 
\mathcal{N}_e \equiv \int_{\phi_i}^{\phi_e} \frac{d \phi}{\mpl \sqrt{2 \epsilon(\phi)}} \sim \frac{1}{\eta} \int_{\theta_i}^{\theta_e} \frac{d\theta}{F^\prime(\theta)} \sim \frac{\mathcal{O}(1)}{\eta}.
\end{split}
\end{equation}
The last relation above involving the number of observable e-foldings $\mathcal{N}_e$ uses the notation $\theta \equiv \phi/f$. 
First line of Eq.~\eqref{eq:slow_roll_inflation_param} shows that quite generically the slow-roll parameter $\epsilon$ is parametrically suppressed compared to $\eta$ (for $f \ll \mpl$), thereby naturally explaining the smallness of the yet-unobserved primordial tensor fluctuations \cite{Planck2018Inflation}. 
The observables---spectral tilt of the primordial scalar fluctuations ($1 - n_s$), tensor-to-scalar ratio ($r$), and the scalar power spectrum amplitude ($A_s$)---as per the Planck CMB data \cite{Planck2018CosmoParam, Planck2018Inflation} are
	\begin{equation}
		\label{eq:obs_slow_roll_param}
	\begin{split}
	1 - n_s = 6 \epsilon - 2 \eta \approx - 2 \eta \approx 0.04 \ , \ 
	r = 16 \epsilon < 0.06 \ , \ 
	A_s \approx 2\times10^{-9},
	\end{split}
	\end{equation}
where, in the first part above, we assume $\epsilon \ll \eta$ as is the case preferred by the data.
Also, as the spectral tilt constraint above shows, $\eta < 0$ is strongly preferred, especially for the low-scale models we are considering (i.e. for small $\epsilon$). A convex form of $F\left(\frac{\phi}{f}\right)$ in Eq.~\eqref{eq:effective_single_field_inflation}, or more generally convex $v(\phi)$ in Eq.~\eqref{eq:effective_inf_potential_Linde}, e.g. $v(\phi) = + \frac{1}{2} m_\phi^2 \phi^2$ as mentioned earlier, gives $\eta > 0$ and hence a blue spectral tilt which is strongly disfavored. Hence, we consider a hilltop-like $F\left(\frac{\phi}{f}\right)$ with inflation happening somewhat close to its maximum. 
Eq.~\eqref{eq:obs_slow_roll_param} constrains the 
parameters of the effective single-field inflation as described by Eq.~\eqref{eq:effective_single_field_inflation}, i.e. $(\vinf, V_0, f)$, as\footnote{We will do a better job of estimating these parameters, especially $\frac{f}{H}$, in Sec.~\ref{subsec:U(1)_UV_completion}, taking the $\sim \mathcal{O}(1)$ factors in $F$ and its derivatives from Eq.~\eqref{eq:effective_single_field_inflation} into account.}
\begin{equation}
\label{eq:effective_hybrid_inflation_param}
\frac{f}{H} \sim \frac{0.1}{\eta \sqrt{A_s}} \sim 10^{6} \ , \
\frac{V_0}{f^4} \sim 10^2 \eta^3 A_s \sim 10^{-12} \ , \
\frac{V_0}{\vinf} \sim \frac{\epsilon}{\eta} \sim \mathcal{O}(10) \ r.
\end{equation}

Hilltop inflation models, in order to satisfy the slow roll conditions, typically require inflation to happen very close to the hilltop. However, with a large offset in the vacuum energy as in Eq.~\eqref{eq:effective_single_field_inflation}, this tuning in the initial inflaton field location is not required. Here, the potential generically satisfies slow-roll conditions for all values of $\phi$ and not just near its extrema. As can be seen in Eq.~\eqref{eq:slow_roll_inflation_param}, $\mathcal{N}_e \propto 1/\eta \sim \mathcal{O}(100)$. Hence, the dimensionless integral there needs only to be $\mathcal{O}(1)$ to get $\mathcal{N}_e = 50-60$ which can be easily satisfied with $\phi_i , \phi_e \sim \mathcal{O}(f)$.


\section{Hybrid ``Twinflation''}
\label{sec:the_model}

In the present section, we propose a natural model for hybrid inflation, ``Twinflation'', which satisfies naturalness for all parameters, all mass scales and field values being smaller than the respective UV cutoff scales, and sub-Planckian physics. We will also make use of the estimates in Sec.~\ref{sec:effective_single_field_inflation}, since the effective inflationary potential here has the same form as in Eq.~\eqref{eq:effective_single_field_inflation}, as we will see later. 

In order to get rid of the quadratic sensitivity of the inflaton potential $V_{\rm eff}(\phi)$ towards the UV physics,
we consider mirroring the $\sigma$-field with a $\mathbb{Z}_2$ exchange symmetry. Considering the original structure of hybrid inflation, Eq.~\eqref{eq:basic_hybrid_inflation_on_hilltop}, one could try $g \phi^2 \sigma^2 \rightarrow g \phi^2 \left(\sigma_A^2  - \sigma_B^2\right)$, such that the quadratic sensitivity of the inflaton mass to the UV scale is canceled between $\sigma_A$ and $\sigma_B$. However, no symmetry protects this structure and hence it is not radiatively stable.
Instead, we consider twinning the $\sigma$-field in our variant hybrid inflation, Eq.~\eqref{eq:soft_coupling_hybrid_inflation}, i.e.
\begin{equation}
	\mu \phi \sigma^2 \rightarrow \mu \phi \left(\sigma_A^2 - \sigma_B^2 \right).
\end{equation}
Here, $m_\phi^2$ has already only log-sensitivity to the UV scale. Now the twinning in $\sigma$ prevents a quadratically divergent $\phi$-tadpole, and thereby removing the associated issues as discussed in Sec.~\ref{sec:soft_coupling}. Also, there exists a symmetry protecting this structure: $\sigma_A \rightarrow \sigma_B , \phi \rightarrow -\phi$; along with $\sigma$-parity i.e. $\sigma_i \rightarrow - \sigma_i$ ($i = A, B$) for simplicity.\footnote{In the next section we will softly break the $\sigma-$parity in a controlled manner to address the cosmological domain wall problem while ensuring naturalness.}  So, this structure is radiatively stable. This can also be realized by a UV completion where $\phi$ is a pNGB of a $U(1)$ global symmetry with soft explicit breaking (see Sec.~\ref{subsec:U(1)_UV_completion}).

A similar model construction to the one presented in the Sec.~\ref{subsec:basic_model}, i.e. Eqs.~\eqref{eq:symmstruc} and \eqref{eq:basic_model_with_renormalizable_terms}, was considered in Ref.~\cite{Berezhiani:1995am} but in the context of mirror-world models to achieve asymmetric reheating of the mirror sector so as to avoid the $\Delta N_{\rm eff}$ constraints. However, here our primary goal is to point out the utility of the twin symmetry in Eq.~\eqref{eq:symmstruc} to address the $\eta-$problem for the inflaton, by constraining inflaton radiative corrections, while reheating can proceed as in standard hybrid inflation.

\subsection{Basic model}
\label{subsec:basic_model}

We now consider the symmetry structure described above, namely,
\begin{equation}\label{eq:symmstruc}
	\sigma_A \rightarrow \sigma_B \ , \ \phi \rightarrow -\phi 
\end{equation}
under the twin symmetry, and also $\sigma_i \rightarrow - \sigma_i$ for simplicity. The most general potential consistent with the above symmetry is given by
\begin{equation}
\label{eq:basic_model_with_renormalizable_terms}
\begin{split}
V(\phi, \sigma_{A,B}) =~ 
&\vinf + \lb - \frac{1}{2}m_\phi^2 \phi^2 + \frac{\lambda_\phi}{4} \phi^4 + \dots \rb \\
&+ \left(\left(\frac{1}{2}M_\sigma^2 \sigma_{A}^2 + \frac{\lambda_\sigma}{4} \sigma_{A}^4\right) + (A \rightarrow  B) \right) + \frac{\bar{\lambda}_\sigma}{4} \sigma_{A}^2 \sigma_{B}^2 \\
&+ \frac{\mu}{2} \phi \left(\sigma_A^2 - \sigma_B^2 \right)
+ \kappa \phi^2 \left(\sigma_A^2 + \sigma_B^2\right) + \dots ,
\end{split}
\end{equation}
where ellipsis after the last term includes higher-dimensional interaction terms, as in Eq.~\eqref{eq:soft_coupling_hybrid_inflation}.
Approximate shift symmetry for the inflaton $\phi$ then requires
\begin{equation}
	\mu , m_\phi \ll M_\sigma \ \ \textrm{and} \ \ \kappa , \lambda_\phi \ll \lambda_\sigma , \bar{\lambda}_\sigma ~~ ,
\end{equation}
which ensures that $\phi$ is much lighter and weakly coupled as compared to $\sigma_i$.

Let us first analyze the effective inflationary dynamics at tree-level. During inflation, i.e. for $\mu \phi < M_\sigma^2$, both the $\sigma$ fields remain heavy and with vanishing VEVs.
Then, integrating them out at tree-level is simply dropping $\sigma_i$ in Eq.~\eqref{eq:basic_model_with_renormalizable_terms}. This gives
\begin{equation}
\label{eq:tree-level_Veff}
	V_{\rm{eff}}(\phi) 
	= \vinf + \lb - \frac{1}{2}m_\phi^2 \phi^2 + \frac{\lambda_\phi}{4} \phi^4 + \dots \rb
	= \vinf + \frac{\lambda_\phi}{4} (\phi^2 - f^2)^2 + \dots ,
\end{equation}
where $f \sim m_\phi / \sqrt{\lambda_\phi}$ and the ellipsis includes sub-dominant higher-dimensional terms in $\phi$. This potential is of the form of Eq.~\eqref{eq:effective_single_field_inflation} and hence all the results of Sec.~\ref{sec:effective_single_field_inflation}, in particular Eq.~\eqref{eq:effective_hybrid_inflation_param}, apply here.
We will consider inflationary trajectory somewhat close to the hilltop of $V_{\rm{eff}}(\phi)$ (i.e. $\phi = 0$), but still with a typical inflaton field value of $\sim \mathcal{O}(f)$ to avoid any considerable initial location tuning. 
As $\phi$ rolls down its potential, $M_{\sigma_i}^2$ change as 
\begin{equation}
	M^2_{\sigma_{A,B}}(\phi) = M^2_\sigma \pm \mu \phi.
\end{equation} 
In order for the waterfall effect to take place, we need 
\begin{equation}
\label{eq:msig}
	M_\sigma^2 \sim \mathcal{O}(\mu f).
\end{equation}
Since $M_{\sigma_A}^2$ always stays positive along the inflationary trajectory, $\sigma_{A}$ has no dynamical role in the model. But $\sigma_B$, which is the true waterfall field here, turns tachyonic at $\phi_* = \frac{M_\sigma^2}{\mu} \sim \mathcal{O}(f)$ and rapidly rolls down to its new minimum. 
The global minimum can be characterized by 
\begin{equation}
	~ \sigma_{B, \textrm{min}} = \lb\frac{\mu \phi_{\rm min} - M_\sigma^2}{\lambda_\sigma}\rb^{1/2} = \frac{M_\sigma}{\sqrt{\lambda_\sigma}} \lb\frac{\phi_{\rm min}}{\phi_*} - 1\rb^{1/2} ,
	~ \sigma_{A, \textrm{min}} = 0 .
\end{equation}
This fast rolling to the global minimum ends inflation by releasing the vacuum energy given by
\begin{equation}
\label{eq:Vinf_from_sigma_sector_vac_energy}
	V_{\rm inf} = \frac{M_\sigma^4}{4\lambda_\sigma} \lb \frac{\phi_{\rm min}}{\phi_*} - 1\rb^2 \sim \mathcal{O}(1) \frac{\mu^2 f^2}{\lambda_\sigma}.
\end{equation} 
In the last step above, as also alluded to before in Sec.~\ref{sec:soft_coupling}, we have set $\phi_{\rm min} \sim \mathcal{O}(\phi_*) \sim \mathcal{O}(f)$ assuming that the higher-dimensional interaction terms in the ellipsis in Eq.~\eqref{eq:basic_model_with_renormalizable_terms} fix the global minimum in $\phi$ at $\sim \mathcal{O}(f)$. As we will see later in Sec.~\ref{subsec:U(1)_UV_completion}, this can be easily realized in a more complete model with $\phi$ as pNGB of a $U(1)$ global symmetry. 


\subsection{Radiative stability and naturalness}
\label{subsec:radiative_stability_naturalness}

In order for the tree-level analysis of the Twinflation model from the previous section to be valid even at loop-level, we need the radiative corrections in Eq.~\eqref{eq:basic_model_with_renormalizable_terms} to be sufficiently small which we explore in this section.
The effect of loops is two-fold: renormalizing tree-level parameters, and giving non-analytic field-dependence via logarithmic terms in the Coleman-Weinberg (CW) potential. First, we require that renormalization of tree-level parameters respects radiative stability and naturalness, and get the resulting constraints on the model parameters. 
Then, in Sec.~\ref{subsec:CW_potential}, we also consider the effects of the full CW potential, but we will show that they can have significant effects only at the boundary of the allowed parameter space, i.e. when naturalness in $V_{\rm eff}(\phi)$ is saturated, which we examine numerically and show in Fig.~\ref{fig:param_space_plot}. In this section, we will therefore defer the full CW analysis in order to first identify the bulk of the viable parameter space.

Here we look for the  constraints in the parameter space required to achieve naturalness of the tree-level parameters. In the $\sigma$-sector, quadratic divergence in $M^2_\sigma$ is induced by the $\sigma$ self-quartic couplings as
\begin{equation}
\delta M^2_{\sigma,\rm 1-loop} \sim \frac{\lambda_\sigma \Lambda_\sigma^2}{16 \pi^2} +  \frac{\bar{\lambda}_\sigma \Lambda_\sigma^2}{16 \pi^2} .
\end{equation}
Hence, naturalness in $M^2_\sigma$ demands the cutoff in $\sigma$-sector to be 
\begin{equation}
\label{eq:sigma_cutoff}
\frac{M_\sigma}{\sqrt{\lambda_\sigma}} \lesssim \Lambda_\sigma \lesssim 4 \pi \frac{M_\sigma}{\sqrt{\lambda_\sigma}}.
\end{equation}
The first constraint above is obtained by demanding that the VEV of $\sigma$ is smaller than the UV scale, which is one of our EFT consistency requirement. We also consider $\bar{\lambda}_\sigma \lesssim \lambda_\sigma$ such that the upper bound on $\Lambda_\sigma$ is controlled by $\lambda_\sigma$ as above. Since both $\bar{\lambda}_\sigma$ and $\lambda_\sigma$ get the same radiative contributions as mentioned below in Eq.~\eqref{eq:loop_contributions}, this is justified. 

In the $\phi$-sector, for simplicity, first we consider an exact shift symmetry, which is then only softly broken by the $\mu$ term in Eq.~\eqref{eq:basic_model_with_renormalizable_terms}.
Then, the loop-level one-particle irreducible (1PI) effective potential has contributions as follows (here we track only the $\mu$-dependent corrections):
\begin{equation}
	\label{eq:loop_contributions}
\begin{split}
	&\delta m^2_{\phi,\rm 1-loop} \sim \frac{\mu^2}{16 \pi^2} \ln \Lambda_\sigma ~ , \\
	&\delta \lb\lambda_\phi, \lambda_\sigma, \bar{\lambda}_\sigma\rb_{\rm{1-loop}} \sim \frac{\mu^4}{16 \pi^2 M_\sigma^4} \sim \frac{\mu^2}{16 \pi^2 f^2} ~ , \\
	&\delta \kappa_{\rm 1-loop} \sim \frac{\lambda_\sigma \mu^2}{16 \pi^2 M_\sigma^2} \sim \frac{\lambda_\sigma \mu}{16 \pi^2 f} ~.
\end{split}
\end{equation}
Here, we first note that there is no quadratic sensitivity to the UV cutoff scales as in Eq.~\eqref{eq:quadratically_divergent_tadpole}, due to cancellations induced by the twin symmetry, and only a log-sensitivity in $m^2_{\phi}$. 
Now, we will consider even tree-level hard breaking of $\phi$-shift symmetry, i.e. tree-level $\lambda_\phi$ and $\kappa$ couplings, which are comparable to the loop contributions above. 
We will take tree-level values for the other parameters to be at least comparable or bigger than their loop contributions. This gives
\begin{equation}
\label{eq:natural_tree_vertices}
	m^2_{\phi,\rm tree} \gtrsim \frac{\mu^2}{16 \pi^2} ~~ , ~~ 
	\lb\lambda_\sigma, \bar{\lambda}_\sigma\rb_{\rm tree} \gtrsim \frac{\mu^2}{16 \pi^2 f^2} ~~, ~~
	\lambda_{\phi, \rm tree} \sim \frac{\mu^2}{16 \pi^2 f^2} ~~, ~~
	\kappa_{\rm tree} \sim \frac{\lambda_\sigma \mu}{16 \pi^2 f} ~~,
\end{equation}
taking $\ln \Lambda_\sigma \sim \mathcal{O}(1)$.
We note that with the above choice for $m_\phi^2$ and $\lambda_\phi$, the $\phi$-transit scale is indeed $\mathcal{O}(f)$.
But, the tree-level $\lambda_\phi$ and $\kappa$ hard breaking terms now induce quadratic UV-sensitivity in $V_{\rm eff}(\phi)$. 
However, their values satisfying the above constraints are sufficiently small so that naturalness in $m^2_\phi$ can still be maintained as below:
\begin{equation}
\begin{split}
&\delta m^2_{\phi, \rm{1-loop}, (\lambda_\phi)} \sim \frac{\lambda_\phi \Lambda_\phi^2}{16 \pi^2} \sim \frac{\mu^2}{16 \pi^2} \frac{\Lambda_\phi^2}{16 \pi^2 f^2} \lesssim \frac{\mu^2}{16 \pi^2} \lesssim m^2_{\phi,\rm tree} \ , \\
&\delta m^2_{\phi, \rm{1-loop}, (\kappa)} \sim \frac{\kappa \Lambda_\sigma^2}{16 \pi^2} \sim \frac{\mu^2}{16 \pi^2} \frac{\Lambda_\sigma^2}{16 \pi^2 M_\sigma^2 / \lambda_\sigma} \lesssim \frac{\mu^2}{16 \pi^2} \lesssim m^2_{\phi,\rm tree} .
\end{split}
\end{equation}
As can be seen above, this requires cutoffs in the two sectors to be bounded as
\begin{equation}
	\label{eq:cutoffs_in_phi_and_sigma}
	\Lambda_\phi \lesssim 4 \pi f ~~ , ~~ \Lambda_\sigma \lesssim 4 \pi \frac{M_\sigma}{\sqrt{\lambda_\sigma}} ~~ , 
\end{equation}
where the $\sigma$-cutoff also satisfies Eq.~\eqref{eq:sigma_cutoff}. We note that these cutoffs can still be bigger than the respective field values.

\paragraph{Getting a consistent inflationary model: \\ }

In order to get a consistent single-field inflation model, we need to satisfy
\begin{equation}
\label{eq:getting_consistent_inflation}
	m^2_{\phi} \sim \eta H^2 \ , \ 
	M_\sigma \gtrsim H \ , \ 
	\vinf \sim H^2 \mpl^2 \sim \frac{M_\sigma^4}{\lambda_\sigma}.
\end{equation}
The first condition above, along with Eq.~\eqref{eq:natural_tree_vertices}, requires $\mu \lesssim \mathcal{O}(H)$. The second condition, i.e. the $\sigma$ fields being at least heavier than the Hubble scale, combined with $M_\sigma^2 \sim \mu f$ (see Eq.~\eqref{eq:msig}) and $f \sim 10^6 H$ (see Eq.~\eqref{eq:effective_hybrid_inflation_param}), requires $\mu \gtrsim 10^{-6} H$. Together, these constrain the model parameter $\mu$ as 
\begin{equation}
	\label{eq:mu_parameter_constraints}
10^{-6} \lesssim \frac{\mu}{H} \lesssim \mathcal{O}(1).
\end{equation}
The lower bound on $\mu$ above also satisfies $\langle\sigma\rangle \lesssim \mpl$ following Eq.~\eqref{eq:Vinf_from_sigma_sector_vac_energy} and Eq.~\eqref{eq:sigma_cutoff}. A stronger requirement of $\Lambda_\sigma \sim 4 \pi \langle\sigma\rangle \lesssim \mpl$ implies $\frac{\mu}{H} \gtrsim 10^{-3}$.

\paragraph{Lower bound on the Hubble scale:\\}

The third condition in Eq.~\eqref{eq:getting_consistent_inflation}, which relates the inflationary Hubble scale to the model parameters, implies
\begin{equation}
\label{eq:lambda_sigma_tree_value}
\lambda_{\sigma} \sim \frac{M_\sigma^4}{H^2 \mpl^2} \sim \frac{\mu^2 f^2}{H^2 \mpl^2} \sim 10^{22} \frac{\mu^2}{f^2} \frac{H^2}{\mpl^2},
\end{equation}
using Eq.~\eqref{eq:effective_hybrid_inflation_param} in the last step. 
Hence naturalness in $\lambda_\sigma$, i.e. $\lambda_\sigma \gtrsim \frac{\mu^2}{16 \pi^2 f^2}$ (see Eq.~\eqref{eq:natural_tree_vertices}), combined with Eq.~\eqref{eq:lambda_sigma_tree_value} gives a lower bound on the inflationary Hubble scale within our Twinflation model as 
\begin{equation}
	\label{eq:lower_bound_on_H}
H \gtrsim 10^6 \textrm{GeV} .
\end{equation}
This also implies a lower bound on the tensor-to-scalar ratio as $r \gtrsim 10^{-16}$.

As we can see above, naturalness in $\lambda_\sigma$ also implies $H^2\mpl^2 \lesssim 16\pi^2 f^4$ i.e. $\vinf \lesssim \Lambda_\phi^4$, with the $\phi$-cutoff $\Lambda_\phi \lesssim 4 \pi f$. Also, perturbativity of $\lambda_\sigma$ combined with Eq.~\eqref{eq:Vinf_from_sigma_sector_vac_energy} and \eqref{eq:sigma_cutoff} implies $\vinf \lesssim \Lambda_\sigma^4$. Thus, the inflationary energy scale being smaller than the UV scales ensures good EFT control in this model. 

Thus, our Twinflation model of Eq.~\eqref{eq:basic_model_with_renormalizable_terms}, with the parameters satisfying the constraints in Eq.~\eqref{eq:natural_tree_vertices}, exhibits naturalness and EFT control. All the mass scales and the field values are less than the corresponding UV cutoff scales, especially $f \lesssim \Lambda_\phi$ and $\langle \sigma \rangle \lesssim \Lambda_\sigma$. As we will see later in Sec.~\ref{subsec:U(1)_UV_completion}, there is a significant parameter space available satisfying $\Lambda_\phi, \Lambda_\sigma \lesssim \mpl$ (see Fig.~\ref{fig:param_space_plot}) such that we have a truly low-scale, sub-Planckian hybrid inflation model under EFT control, satisfying all of our naturalness requirements as mentioned in Sec.~\ref{sec:hybrid_inflation_review}.


\subsection{One-loop Coleman-Weinberg effective potential}
\label{subsec:CW_potential}

As we noted earlier, the $\sigma$ fields are always heavy before the end of inflation, and hence can be integrated out to give a
1-loop Coleman-Weinberg (CW) potential: 
\begin{equation}
\label{eq:Coleman_Weinberg_potential_basic_model}
\begin{split}
    V_{\rm CW}(\phi) &= \sum_{i=A,B} \frac{M_{\sigma_i}^4(\phi)}{64 \pi^2} \ln{\frac{M_{\sigma_i}^2(\phi)}{\Lambda_\sigma^2}} \\
    &= \frac{\mu^2 f^2}{64 \pi^2} \left[ \lb 2 \frac{\phi^2}{f^2} + \cdots \rb \ln{\frac{\mu f}{\Lambda_\sigma^2}} 
    +  \frac{(\phi_* +\phi)^2}{f^2} \ln{\frac{\phi_* +\phi}{f}}
    + \frac{(\phi_* - \phi)^2}{f^2} \ln{\frac{\phi_* - \phi}{f}} \right] .
\end{split}
\end{equation}
The first term above renormalizes $m_{\phi,\text{tree}}^2$ as in Eq.~\eqref{eq:loop_contributions}. 
Parameterizing the tree-level inflaton mass as
\begin{equation}
m_{\phi,\rm {tree}}^2 \equiv c_\phi \frac{\mu^2}{16\pi^2} \ ,
\end{equation}
the naturalness constraint in Eq.~\eqref{eq:natural_tree_vertices} requires $c_\phi \gtrsim \mathcal{O}(1)$. 
Then, $V_{\rm CW} (\phi)$ in Eq.~\eqref{eq:Coleman_Weinberg_potential_basic_model} is comparable to tree-level $V_{\rm eff} (\phi)$ in Eq.~\eqref{eq:tree-level_Veff} only when $c_\phi \approx 1$, while giving sub-dominant effects for the bulk of the natural parameter space ($c_\phi \gg 1$).
Nevertheless, in our full numerical analysis in Sec.~\ref{subsec:U(1)_UV_completion}, we will incorporate the logarithmic effects in the inflaton that distinguish the 1-loop potential, but they are so modest as to be difficult to resolve by eye,  
as we will see in Fig.~\ref{fig:param_space_plot}.


\subsection{Pseudo-Nambu-Goldstone inflaton realization}
\label{subsec:U(1)_UV_completion}

In this section, we discuss a simple and more  complete extension  of the model in Eq.~\eqref{eq:basic_model_with_renormalizable_terms}, realizing the inflaton as a pNGB of a 
global $U(1)$ symmetry, with soft explicit breaking. The Lagrangian is given by,
\begin{align}
	\label{eq:UVmodel}
\mathcal{L}_{\rm UV} = &|\partial\Phi|^2-V_\Phi(|\Phi|^2)\nonumber\\
&+\lb\lb\frac{1}{2}(\partial\sigma_A)^2-\frac{1}{2}M_\sigma^2 \sigma_{A}^2 - \frac{\lambda_\sigma}{4} \sigma_{A}^4\rb + (A \rightarrow  B) \rb - \frac{\bar{\lambda}_\sigma}{4} \sigma_{A}^2 \sigma_{B}^2 \nonumber\\
&+\lb\frac{\mu\Phi}{2 \sqrt{2}}(\sigma_{A}^2-\sigma_B^2) + \frac{c_\phi}{64 \pi^2}\lb\mu\Phi\rb^2+\text{h.c.}\rb - g|\Phi|^2\lb\sigma_A^2+\sigma_B^2\rb - \vinf.
\end{align}
Similar to the symmetry structure in Eq.~\eqref{eq:symmstruc}, we demand
\begin{align}
\label{eq:symmetry_in_U(1)_model}
\Phi\rightarrow-\Phi, ~~\sigma_A\rightarrow\sigma_B
\end{align}
under the twin symmetry, and also for simplicity a $\mathbb{Z}_2$-symmetry under which $\sigma_i \rightarrow -\sigma_i$ for $i=A,B$. Furthermore, we treat $\mu$ as a $U(1)$ ``spurion'' with charge $-1$ that compensates the $+1$ charge of $\Phi$ under the $U(1)$. This spurion analysis, along with the symmetry structure in Eq.~\eqref{eq:symmetry_in_U(1)_model}, uniquely fixes the Lagrangian in Eq.~\eqref{eq:UVmodel} at the dimension-4 level. There are two dimensionless coupling constants $c_\phi$ and $g$, with $\mu,M_\sigma,\lambda_\sigma,\bar{\lambda}_\sigma$ being the same as in Eq.~\eqref{eq:basic_model_with_renormalizable_terms}.\footnote{To simplify the notation, we keep using the same parameter $\mu$ as before, although now it has a spurion charge.} The potential $V_\Phi$ is such that it allows for a spontaneous breaking of $U(1)$ with the inflaton ($\phi$) being the corresponding Nambu-Goldstone boson (NGB). The $\mu-$term in the third line of Eq.~\eqref{eq:UVmodel} then gives mass to the inflaton, as we will see below, making it a pseudo-NGB. We parametrize the inflaton $\phi$ as $\Phi=\frac{f+\chi}{\sqrt{2}}e^{i\phi/f}$, where $\chi$ is the radial mode and $\langle\Phi\rangle = f$ is the VEV. Integrating out $\chi$ and redefining $\frac{\phi}{f} \rightarrow \frac{\phi}{f} + \pi/2$, we get an effective Lagrangian from Eq.~\eqref{eq:UVmodel} as
\begin{align}\label{eq:IRmodel}
\mathcal{L}_{\rm IR}=
&\lb\lb\frac{1}{2}(\partial\sigma_A)^2-\frac{1}{2}\widetilde{M}_\sigma^2 \sigma_{A}^2 - \frac{\lambda_\sigma}{4} \sigma_{A}^4\rb + (A \rightarrow  B) \rb - \frac{\bar{\lambda}_\sigma}{4} \sigma_{A}^2 \sigma_{B}^2 \nonumber\\
&+\frac{1}{2}(\partial \phi)^2 - \frac{\mu f}{2} \sin\lb\frac{\phi}{f}\rb\lb\sigma_A^2-\sigma_B^2\rb - c_\phi \frac{\mu^2 f^2}{64 \pi^2} \cos\lb\frac{2\phi}{f}\rb - \vinf.
\end{align}
Here we have defined $\widetilde{M}_\sigma^2 \equiv M_\sigma^2 + g f^2$.
For the waterfall mechanism to work, we need 
both $M_\sigma^2\sim \mu f$, which was discussed earlier, and $g \lesssim \mu/f$, which then implies $\widetilde{M}_\sigma^2\sim M_\sigma^2 \sim \mu f$. Hence, in what follows, we will drop the tilde over $M_\sigma^2$. This value of $g$ is technically natural since loop-contributions in the 1PI effective potential include
\begin{align}
\delta g_{\rm 1-loop} \sim \frac{\lambda_\sigma \mu^2}{16 \pi^2 M_\sigma^2} \sim \frac{\lambda_\sigma\mu}{16\pi^2f} \ll \frac{\mu}{f}. 
\end{align}
Inflation starts somewhat near the hilltop along $\phi$ i.e. close to $\phi = 0$. 
Expanding for $\phi/f\ll 1$ in Eq.~\eqref{eq:IRmodel}, we get\footnote{The size of the cosine potential in $\phi$ ($\sim \mu^2 f^2/16 \pi^2$) is much smaller than $\vinf \sim  \mu^2 f^2/\lambda_\sigma$, as we will see later in Eq.~\eqref{eq:Vinf_U(1)_model}, and hence the constant term from the cosine can be neglected here.}
\begin{align}
\mathcal{L}_{\rm IR}\approx
&\lb\lb\frac{1}{2}(\partial\sigma_A)^2-\frac{1}{2} M_\sigma^2 \sigma_{A}^2 - \frac{\lambda_\sigma}{4} \sigma_{A}^4\rb + (A \rightarrow  B) \rb - \frac{\bar{\lambda}_\sigma}{4} \sigma_{A}^2 \sigma_{B}^2 \nonumber\\
&+\frac{1}{2}(\partial \phi)^2 - \frac{\mu \phi}{2} \lb\sigma_A^2-\sigma_B^2\rb - \vinf
+ c_\phi \frac{\mu^2}{16 \pi^2} \lb \frac{\phi^2}{2} - \frac{\phi^4}{6 f^2} + \dots \rb.
\end{align}
For $ c_\phi \gtrsim \mathcal{O}(1)$, as required by technical naturalness in Eq.~\eqref{eq:UVmodel}, this reproduces all the interactions relevant for hybrid inflation as was studied earlier in Eq.~\eqref{eq:basic_model_with_renormalizable_terms} for $c_\phi>0$. 

During inflation, i.e. with $\sin \lb\frac{\phi}{f}\rb < \frac{M_\sigma^2}{\mu f}$, both $\sigma_{A,B}$ remain heavy and with vanishing VEVs. Thus, integrating them out at tree-level, which is dropping them in Eq.~\eqref{eq:IRmodel}, gives an effective inflationary potential
\begin{equation}
	\label{eq:effective_inflationary_potential_from_U(1)_model}
	V_{\rm eff}(\phi) \approx \vinf + c_\phi \frac{\mu^2 f^2}{64 \pi^2}  \cos\lb\frac{2 \phi}{f} \rb.
\end{equation}
This is of the form of Eq.~\eqref{eq:effective_single_field_inflation} with the function $F\lb\frac{\phi}{f}\rb$ taking trigonometric form as above, and hence all the results of Sec.~\ref{sec:effective_single_field_inflation} apply here too. 
As inflaton rolls past a critical value $\phi_*$ such that 
\begin{equation}
	\label{eq:phi_*_U(1)_model}
	\sin \lb\frac{\phi_*}{f}\rb = \frac{M_\sigma^2}{\mu f},
\end{equation} 
waterfall is triggered along $\sigma_B$. The fields then rapidly roll down to the global minimum which is situated at
\begin{equation}
\begin{split}
	&\frac{\phi_{\rm min}}{f} = \frac{\pi}{2}, ~~ \sigma_{A, \textrm{min}} = 0, \\ 
	&\sigma_{B, \textrm{min}} = \sqrt{\frac{1}{\lambda_\sigma} \lb\mu f \sin\lb\frac{\phi_{\rm min}}{f}\rb - M_\sigma^2 \rb} = \sqrt{\frac{\mu f}{\lambda_\sigma} \lb 1 - \sin\lb \frac{\phi_*}{f}\rb \rb} \sim \mathcal{O}(1) \sqrt{\frac{\mu f}{\lambda_\sigma}} .
	\end{split}
\end{equation} 
The inflationary vacuum energy released during this waterfall transition is given by
\begin{equation}
	\label{eq:Vinf_U(1)_model}
	\vinf \approx \frac{\mu^2 f^2}{4 \lambda_\sigma} \lb 1 - \sin\lb \frac{\phi_*}{f}\rb \rb^2 \sim \mathcal{O}(1) \frac{\mu^2 f^2}{\lambda_\sigma}.
\end{equation}
Thus, as mentioned earlier in Sec.~\ref{subsec:basic_model}, once $\phi$ is realized as a pNGB of a $U(1)$ global symmetry as in this section, the global minimum in $\phi$ is fixed only $\sim \mathcal{O}(1)$ away from the critical point triggering waterfall, i.e. $\phi_{\rm min} \sim \mathcal{O}(\phi_*) \sim \mathcal{O}(f)$. Consequently, the parametric dependence of $\vinf$ (and hence $H$) on the model parameters is obtained as in Eq.~\eqref{eq:Vinf_U(1)_model}, which is as expected in Eq.~\eqref{eq:Vinf_from_sigma_sector_vac_energy}.

\begin{figure}[t]
	\centering
	\includegraphics[width=\textwidth]{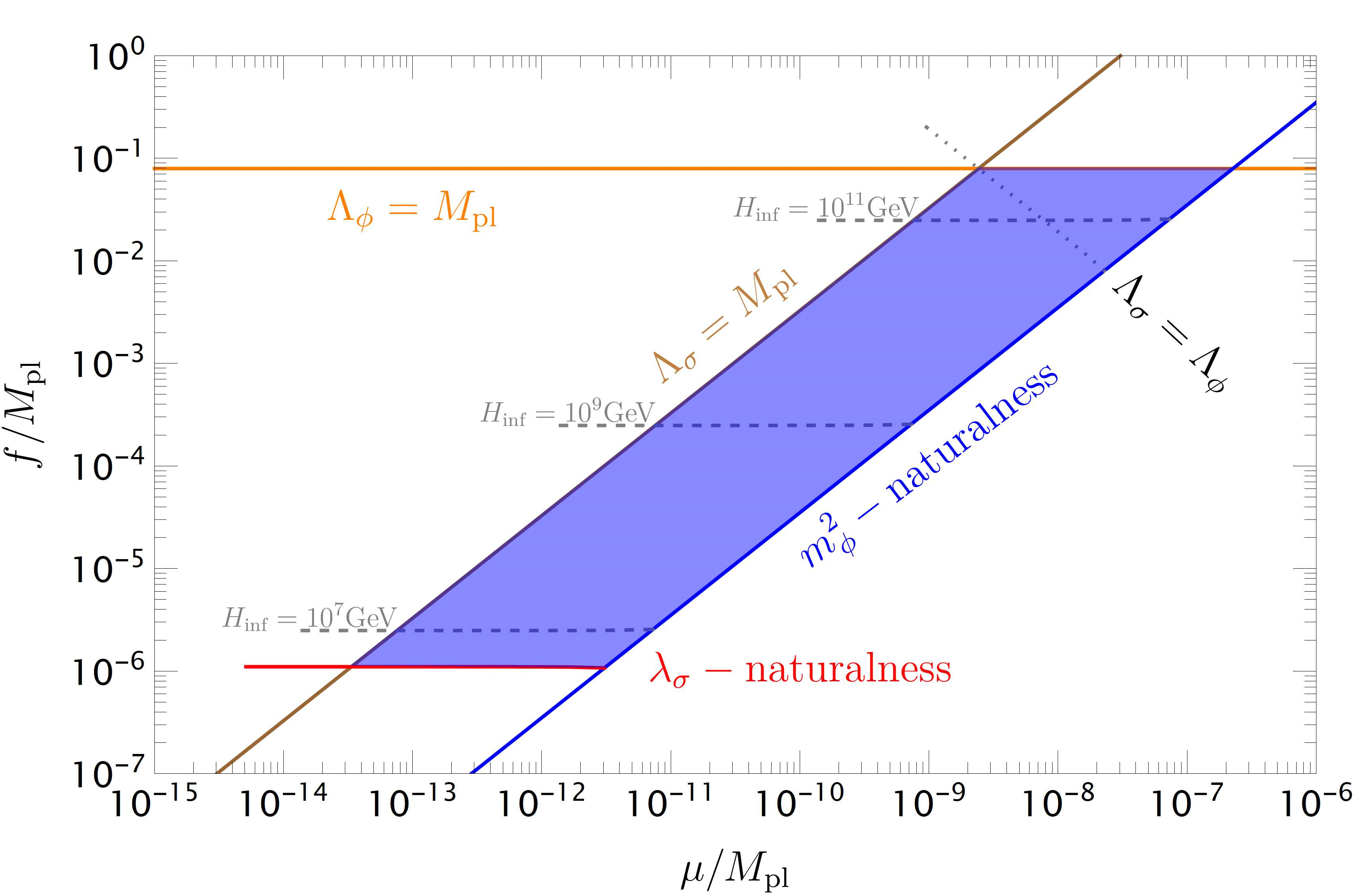}
	\caption{Available parameter space in the $U(1)$ version of our Twinflation model (see Sec.~\ref{subsec:U(1)_UV_completion}) exhibiting naturalness and EFT-control: $\phi_*/f=\pi/5$ for concreteness. The right and bottom edges of the shaded region correspond to naturalness constraints on $m_\phi$ and $\lambda_\sigma$, respectively. The top and left edges correspond to the cutoffs $\Lambda_\phi$ and $\Lambda_\sigma$ being sub-Planckian, respectively. $\Lambda_\phi \approx \Lambda_\sigma$ on the dotted line. The parameter $c_\phi$ varies from 1 to $\sim 10^4$ as we move from right to left edge, which makes the loop contributions to inflaton potential smaller and smaller as compared to the tree-level term. The dashed lines show contours for $H = 10^7, 10^9, 10^{11}$ GeV, corresponding to $r \approx 10^{-15}, 10^{-11}, 10^{-7} $, respectively. $n_s$ is fixed to 0.9649, its central value from the Planck CMB constraints \cite{Planck2018Inflation}. Varying its value up or down by a percent shifts the entire blue region slightly to the left or right, respectively, by about a percent which is hardly resolvable by eye.
	}
	\label{fig:param_space_plot}
\end{figure}

Integrating out the heavy $\sigma$ fields at 1-loop level, similar to Eq.~\eqref{eq:Coleman_Weinberg_potential_basic_model}, gives rise to the following logarithmic dependence from the Coleman-Weinberg potential:
\begin{equation}
\label{eq:Coleman_Weinberg_potential_U(1)_model}
\begin{split}
V_{\rm CW} \lb \theta \equiv \frac{\phi}{f} \rb = \frac{\mu^2 f^2}{64 \pi^2} &\left[\lb \sin \theta_* + \sin \theta\rb^2 \ln{(\sin \theta_* + \sin \theta)} \right. \\  
&+ \left. \lb\sin \theta_* - \sin \theta \rb^2 \ln{(\sin \theta_* - \sin \theta)}  \right]    .
\end{split}
\end{equation}
As mentioned earlier in Sec.~\ref{subsec:radiative_stability_naturalness}, this can give considerable effects only when naturalness is saturated for $m_\phi^2$, i.e. for $c_\phi \approx 1$. These effects, numerically computed in Fig.~\ref{fig:param_space_plot}, 
are however so modest as to be difficult to resolve by eye. 

Fig.~\ref{fig:param_space_plot} shows the available parameter space in our Twinflation model described by Eq.~\eqref{eq:IRmodel}, satisfying the requirements of naturalness and EFT control, and giving a viable hybrid inflation model. 
Here we have fixed $\frac{\phi_*}{f} = \frac{\pi}{5}$ for concreteness. This then gives the initial field value\footnote{This value changes slightly for different $c_\phi$ values, i.e. including the CW potential from Eq.~\eqref{eq:Coleman_Weinberg_potential_U(1)_model}.} $\frac{\phi_i}{f} \approx 0.1 \pi$ to get 60 e-foldings, using the effective potential in Eq.~\eqref{eq:effective_inflationary_potential_from_U(1)_model} and the analysis in Sec.~\ref{sec:effective_single_field_inflation}. 
This gives the trigonometric functions $\sim \mathcal{O}(1)$ for both $\frac{\phi_i}{f}$ and $\frac{\phi_*}{f}$, as alluded to before in Sec.~\ref{sec:effective_single_field_inflation}.
The other essential parameters $M_\sigma^2$ and $\lambda_\sigma$ are then fixed by the model requirements in Eqs.~\eqref{eq:phi_*_U(1)_model}, \eqref{eq:Vinf_U(1)_model}, and \eqref{eq:effective_hybrid_inflation_param}.
The right and bottom edges of the allowed parameter space correspond to naturalness constraints on $m_\phi$ (see Eq.~\eqref{eq:mu_parameter_constraints}) and $\lambda_\sigma$ (see Eq.~\eqref{eq:natural_tree_vertices}), respectively. The top and left edges correspond to the cutoffs in the $\phi$ and $\sigma$ sectors being sub-Planckian, respectively. Here we consider $\Lambda_\phi \approx 4 \pi f , \Lambda_\sigma \approx 4 \pi \frac{M_\sigma}{\sqrt{\lambda_\sigma}}$ saturating the constraints in Eq.~\eqref{eq:cutoffs_in_phi_and_sigma}. Thus, the shaded region satisfies our naturalness and EFT consistency requirements.
$n_s$ is fixed to 0.9649, its central value from the Planck CMB constraints \cite{Planck2018Inflation}. Varying its value up or down by a percent shifts the entire allowed region slightly to the left or right, respectively, by about a percent.
The dashed lines show contours for $H$ which are mostly horizontal (i.e. constant $f/H$, see Eq.~\eqref{eq:effective_hybrid_inflation_param}), but bending slightly upwards close to the right edge due to the CW potential contribution.
As we can see in the figure, $\Lambda_\phi$ being sub-Planckian restricts the model to realize $H \lesssim 10^{11}$ GeV, while the $\lambda_\sigma$-naturalness gives a lower bound on $H$ as $\sim 10^6$ GeV as expected from Eq.~\eqref{eq:lower_bound_on_H}. 
The two cutoffs $\Lambda_\phi , \Lambda_\sigma$ are approximately equal on the dotted line. Thus, as the figure shows, demanding $\Lambda_\phi \approx \Lambda_\sigma$ can  only realize $H$ bigger than $\sim 10^{10}$ GeV. Only a small part of the parameter space lying above this dotted line corresponds to $\Lambda_\phi > \Lambda_\sigma$, while a majority of the allowed region has $\Lambda_\sigma > \Lambda_\phi$.

The Lagrangian of the $U(1)$ model in Eq.~\eqref{eq:UVmodel} contains terms only up to dimension-4. This will also include higher-dimensional terms respecting the symmetry in Eq.~\eqref{eq:symmetry_in_U(1)_model} and the spurion analysis mentioned thereafter, and thus will be of the form
\begin{equation}
	\label{eq:U(1)_higher_dim_terms}
	\delta \mathcal{L}_{\rm UV, non-ren.} \ni c_{nm} \frac{\lb\mu \Phi \rb^n \lb\sigma_i^2\rb^m}{\lb\Lambda^2\rb^{n+m-2}} ~~ .
\end{equation}
Here, the exponents $n, m$ and the combinations of $\sigma_{A,B}$ in $\sigma_i^{2}$ will be such that they respect the symmetry in Eq.~\eqref{eq:symmetry_in_U(1)_model}. Also, for simplicity, we consider here a single UV cutoff scale $\Lambda$ suppressing these non-renormalizable terms.\footnote{It can be shown that even with different cutoff scales for $\phi$ and $\sigma$ fields, analogous to what is shown here for $\Lambda_\phi \sim \Lambda_\sigma$, these non-renormalizable terms do not pose any danger to our model.} 
In order to satisfy naturalness in the $\sigma$-potential, it suffices to have 
$c_{0 m} \lesssim \lb 16 \pi^2 \rb^{m - 2} \lambda_\sigma$. 
This mild requirement on the coefficients $c_{nm}$ in Eq.~\eqref{eq:U(1)_higher_dim_terms}, i.e. $c_{nm} \sim c_{0 m} \lesssim  \lb 16 \pi^2 \rb^{m - 2} \lambda_\sigma$, is sufficient to render the entire model natural, even at the non-renormalizable level, as illustrated below. The most vulnerable terms would be the super-renormalizable terms in Eq.~\eqref{eq:UVmodel}, i.e. the bare and $\Phi-$dependent $\sigma$ mass terms, which we collectively refer to as $M_\sigma^2 (\Phi)$. The higher-dimensional terms in Eq.~\eqref{eq:U(1)_higher_dim_terms} can contribute to $M_\sigma^2 (\Phi)$ at loop- or tree-level (i.e. after setting some fields to their VEVs) as 
\begin{equation}
	\frac{\delta M_\sigma^2 (\Phi)}{M_\sigma^2} 
	\sim  \frac{c_{nm} (\mu \Phi)^n \cdot \langle \sigma \rangle^{2(m-1)}}{M_\sigma^2 \cdot \Lambda^{2 (n+m-2)} } 
	\lesssim \frac{(16 \pi^2)^{m - 2} (\mu \Phi)^n \cdot \langle \sigma \rangle^{2(m-2)}}{\Lambda^{2(n+m-2)}} 
	\sim \lb\frac{\mu \Phi}{\Lambda^2}\rb^n 
	\lesssim \lb\frac{\mu}{\Lambda}\rb^n  ,
\end{equation}
which is negligible due to the suppression from $\frac{\mu}{\Lambda} \lesssim \frac{H}{4 \pi f} \lesssim 10^{-6}$. Also, any higher-dimensional terms in Eq.~\eqref{eq:UVmodel} involving $|\Phi|^2$ will be sub-dominant since they will come with suppression factors of at least $\frac{|\Phi|^2}{\Lambda^2} \sim \frac{1}{16 \pi^2}$. 


\section{Addressing the cosmological domain wall problem}
\label{sec:domain_wall_problem}

Spontaneous breaking of an exact discrete symmetry, in our model $\sigma_i \rightarrow - \sigma_i$, during cosmological evolution, will lead to the formation of domains (with $\langle \sigma_B \rangle > 0$ or $< 0$) after the end of inflation, separated by cosmologically stable domain walls (DW). The energy density in these domain walls redshifts slower than both matter and radiation. 
This gives rise to a late-time universe dominated by domain walls contrary to what is observed during Big-Bang Nucleosynthesis.
This is the so called ``cosmological domain wall problem'' \cite{DomainWallProblem_Zeldovich:1974uw}, which our Twinflation model faces for an exact $\sigma_i \rightarrow - \sigma_i$ symmetry.
The $\sigma$ fields could be charged under a $U(1)$ gauge symmetry, which then may not give rise to domain walls, but instead forms the much less constrained cosmic strings (see e.g. \cite{Vilenkin:1982ks, Hindmarsh:2011qj, Auclair:2019wcv}). However, this approach requires additional fields and structures. 
Here we will consider a simple solution to the domain wall problem via small explicit breaking of the discrete symmetry.

We first note that $\sigma_i \rightarrow - \sigma_i$ symmetry is not an essential ingredient of our model and is used so far only for simplicity. We can hence add a small soft breaking of this symmetry in Eq.~\eqref{eq:basic_model_with_renormalizable_terms} or \eqref{eq:IRmodel} via
\begin{equation}
	\label{eq:cubic_term_in_sigma}
	V(\phi, \sigma_i) \ni M \sigma_i^3 , 
\end{equation}
where $M$ is a dimensionful spurion of this $\sigma$-parity breaking. This leads to a bias between the previously degenerate vacua as
\begin{equation}
	\label{eq:Vbias_from_sigma_cubic}
	\frac{\Delta V_{\rm bias}}{\vinf} \sim \frac{M}{M_\sigma \sqrt{\lambda_\sigma}} , 
\end{equation}
where in the denominator we have $\vinf$ which is also the typical size of the $\sigma$-potential. This bias provides a pressure force acting against the surface tension of the walls, eventually leading to their annihilation. Then, demanding that this annihilation of domain walls happens before their cosmological energy domination, we need \cite{DomainWallBias_Vilenkin:1981zs, DomainWallBias_Gelmini:1988sf, DomainWalls_Saikawa:2017hiv}
\begin{equation}
	\mathcal{O}(1) \gtrsim \frac{\Delta V_{\rm bias}}{\vinf} \gtrsim \frac{M_\sigma^2}{\lambda_\sigma \mpl^2} ,
\end{equation}
which can be realized in our model, using Eq.~\eqref{eq:Vbias_from_sigma_cubic}, by having
\begin{equation}
	\label{eq:M_constraints}
 	M_\sigma \sqrt{\lambda_\sigma} \gtrsim M \gtrsim \frac{M_\sigma^3}{\sqrt{\lambda_\sigma} \mpl^2}.
\end{equation}
However, the cubic term in Eq.~\eqref{eq:cubic_term_in_sigma} radiatively generates the following $\sigma$-tadpole:
\begin{equation}
	V(\phi, \sigma_i) \ni M \frac{\Lambda_\sigma^2}{16 \pi^2} \sigma_i \sim M \frac{M_\sigma^2}{\lambda_\sigma} \sigma_i .
\end{equation}
Tadpole terms of this order shift the minimum in $\sigma_i$ in a $\phi$-dependent way as
\begin{equation}
	\label{eq:sigma_shift_due_to_tadpole}
	\delta \sigma_i (\phi) \sim \frac{M M_\sigma^2}{\lambda_\sigma M_{\sigma_i}^2 (\phi)} 
	\sim \frac{M}{\lambda_\sigma} \lb 1 \pm \frac{\sin(\phi/f)}{\sin(\phi_*/f)} \rb^{-1}, 
\end{equation}
where $M_{\sigma_i}^2(\phi) = M_\sigma^2 \pm \mu f \sin\lb\phi/f\rb$ is the $\phi$-dependent mass-squared for $\sigma_i$  (see Eq.~\eqref{eq:IRmodel}).
This shift contributes to the effective inflaton potential as\footnote{As $\phi \rightarrow \phi_*$, i.e. towards the end of inflation, the expressions in Eqs.~\eqref{eq:sigma_shift_due_to_tadpole}, \eqref{eq:delta_Veff_from_sigma_tadpole} seem to diverge. However, this is because the effective mass for $\sigma_B$ vanishes at $\phi_*$, and hence we have to balance the $\sigma$-tadpole with $\sigma$-cubic which will modify these expressions close to $\phi_*$. 
}
\begin{equation}
	\label{eq:delta_Veff_from_sigma_tadpole}
	\delta V_{\rm eff}(\phi) \sim \sum_{i=A,B} \frac{M^2 M_\sigma^4}{\lambda_\sigma^2 M_{\sigma_i}^2 (\phi)} 
	\sim \frac{M^2 M_\sigma^2}{\lambda_\sigma^2} \lb 1 - \frac{\sin^2(\phi/f)}{\sin^2(\phi_*/f)} \rb^{-1} .
\end{equation}
\begin{figure}[t]
	\centering
	\includegraphics[width=\textwidth]{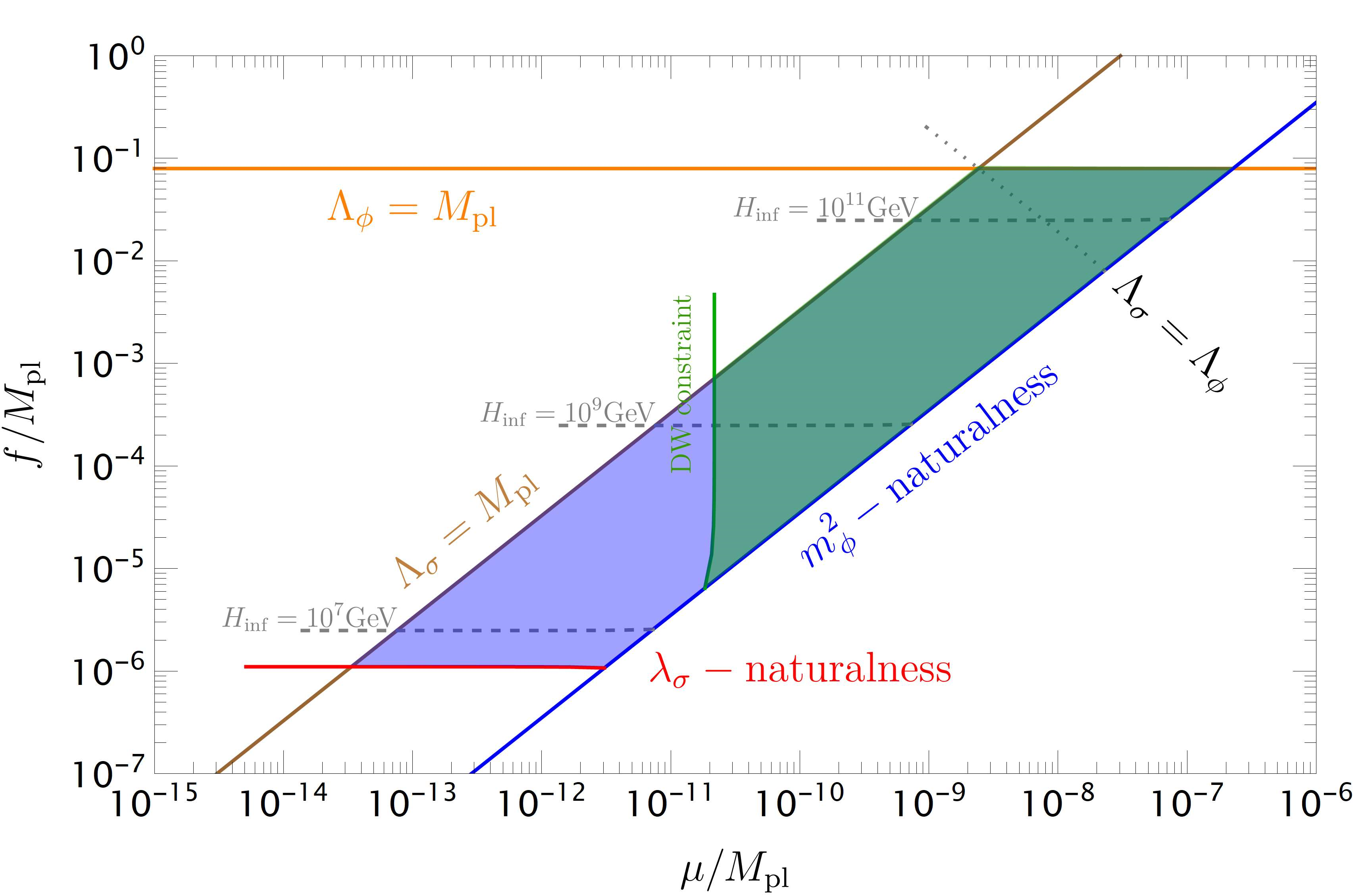}
	\caption{Addressing the cosmological domain wall problem in Twinflation: The blue region (same as in Fig.~\ref{fig:param_space_plot}) satisfies our naturalness and EFT consistency requirements. 
	Small explicit breaking of $\sigma$-parity (see Eq.~\eqref{eq:cubic_term_in_sigma}) solves the domain wall problem. Its contribution to $V_{\rm eff} (\phi)$, via the natural value of  $\sigma$-tadpole, is sub-dominant in the green region shown above. 
	}
	\label{fig:param_space_plot_addressing_domain_wall_issue}
\end{figure}
Demanding that this contribution is sub-dominant to the inflaton potential implies
\begin{equation}
	1 \gtrsim \frac{\delta V_{\rm eff} (\phi)}{V_{\rm eff} (\phi)} \sim \frac{16 \pi^2 M^2}{c_\phi \lambda_\sigma^2 M_\sigma^2} \gtrsim \frac{16 \pi^2 M_\sigma^4}{c_\phi \lambda_\sigma^3 \mpl^4} ,
\end{equation}
where in the last step we have used Eq.~\eqref{eq:M_constraints}.
Then, using our model requirements -- $\lambda_\sigma \sim \frac{M_\sigma^4}{H^2 \mpl^2}, M_\sigma^2 \sim \mu f , \frac{f}{H} \sim 10^6$ -- we get the constraint for the allowed parameter region as
\begin{equation}
	\sqrt{c_\phi} \frac{\mu^2}{f \mpl} \gtrsim 10^{-17} .
\end{equation}
This is evaluated numerically and shown in Fig.~\ref{fig:param_space_plot_addressing_domain_wall_issue} as the 
green region. 
We can also note here that this now gives a lower bound on the Hubble scale as
\begin{equation}
	H \gtrsim 10^7 \textrm{GeV} ,
\end{equation}
which is $\sim \mathcal{O}(10)$ bigger than that obtained in Eq.~\eqref{eq:lower_bound_on_H}.


%
%

Thus, the cosmological domain wall problem can be solved in our model by introducing a small explicit breaking of $\sigma$-parity at the cost of some reduction in the allowed parameter space as shown in Fig.~\ref{fig:param_space_plot_addressing_domain_wall_issue}.
%
%
One might explore more general ways of explicit $\sigma$-parity breaking than the simple one we considered here via Eq.~\eqref{eq:cubic_term_in_sigma}, possibly allowing for viable hybrid inflation in the entire blue region. We leave this exploration for a future study.


\section{Discussion}
\label{sec:discussion}

In the present work, we build a viable, natural, and EFT-controlled model of low-scale hybrid inflation, ``Twinflation''. Here, inflation happens somewhat near the hilltop of the effective inflaton potential, although without any fine-tuning of the initial position. This gives rise to the red tilt in the scalar perturbations, consistent with the observations. The quadratic sensitivity to the UV cutoff scales in the inflaton potential, induced by its necessarily non-derivative coupling with the waterfall field, is removed by a twin symmetry. All the parameters take (technically) natural values, without any fine-tuning. All the mass scales and field values are below the respective UV cutoff scales and also the Planck scale, thus rendering the model under (straightforward) EFT control. This model can realize low-scale inflation with the Hubble scale as low as $\sim 10^6$ GeV (see Fig.~\ref{fig:param_space_plot}). It is therefore easily consistent
with the smallness of the yet-unobserved primordial tensor fluctuations, which could be unobservably small ($ r \sim 10^{-16}$) for the lowest Hubble scales realized in our model.  

Spontaneous breaking of the discrete symmetry $\sigma_i \rightarrow - \sigma_i$ towards the end of inflation will lead to cosmic domain wall formation in the post-inflationary universe. One simple way to be compatible with our universe on the large scales at late times, is to demand that such domain walls should annihilate before they start dominating the cosmic energy density. As discussed in Sec.~\ref{sec:domain_wall_problem}, we show that this can be easily implemented in our model with a small explicit breaking of the $\sigma$-parity, which we only considered for technical simplification in any case. This, however, can be achieved only in the  parameter space as shown in Fig.~\ref{fig:param_space_plot_addressing_domain_wall_issue}, allowing for the smallest inflationary Hubble scale to be $\sim 10^7$ GeV. We expect that allowing for more general ways of explicit $\sigma$-parity breaking can possibly relax this constraint, which we leave for a future study. It is also interesting that the domain wall dynamics can give rise to a stochastic gravitational wave (GW) background observable in future GW experiments. See  \cite{DomainWalls_Saikawa:2017hiv} for a review.

Hybrid inflation models typically require fine-tuned couplings.
However, our model does not require any fine-tuning in the parameters to achieve radiative stability. With regards to the initial conditions, we also showed that there is no tuning required in the initial inflaton field location, i.e. it need not start very close to the hilltop and can have a transit of $\sim \mathcal{O}(f)$. 
A large initial inflaton velocity can be compensated by starting more uphill along the potential, up to the hilltop. However, demanding that it first damps to the terminal slow-roll velocity, then gives the required number of e-foldings of slow-roll inflation before entering the waterfall phase, we see that the initial velocity has to be sufficiently small: $\frac{\dot{\phi}}{f^2} \lesssim \frac{H}{f} \sim 10^{-6}$. (See also \cite{Buchmuller:2014epa} for similar constraints.)
Furthermore, there is the question of whether inflation can begin in an inhomogeneous spacetime. Numerical simulations show that whereas large-field inflation models are less susceptible to inhomogeneities preventing the onset of inflation, small-field inflation models may be more so~\cite{Goldwirth:1991rj,Laguna:1991zs,KurkiSuonio:1993fg,Easther:2014zga,East:2015ggf,Clough:2016ymm}.
These issues can however be addressed, for example, by invoking tunneling from a prior metastable vacuum in the landscape of the theory, which naturally gives rise to a state with small field velocity and inhomogeneity (see e.g. \cite{Freivogel:2005vv, Dutta:2011fe, Guth:2013sya, Masoumi:2017gmh}).


It would obviously be very interesting if we could directly observe the waterfall field(s) ($\sigma_i$) via their mediation of primordial non-Gaussianity (NG), using the idea of ``Cosmological Collider Physics''~\cite{Chen:2009zp,Arkani-Hamed:2015bza}. Ordinarily such signals would be strongly ``Boltzmann''-suppressed by $e^{-\pi M_\sigma/H}$, since $M_\sigma \gg H$. However, the recently discussed ``scalar chemical potential'' mechanism \cite{NG_with_chemical_potential_Bodas:2020yho} may eliminate this suppression and be compatible with our twin symmetry structure. We leave an exploration of this to future work.

As discussed in the Introduction, a variety of UV physics scenarios may give rise to unwanted defects or relics like monopoles, moduli, gravitino (see e.g. \cite{GravitinoProblem_Ellis:1982yb, GravitinoProblem_Ellis:1984eq,  GravitinoProblem_Murayama_etal, ModuliProblem_Randall:1994fr}). Different UV scenarios can also exhibit a meta-stable high temperature phase in which the universe can remain stuck if the phase transition to the familiar low temperature phase fails to complete~\cite{RSPT_Creminelli:2001th}. Reheating of the universe at a low temperature, following inflation with a low Hubble scale, might help to address these issues in a straightforward way.
%
Another motivation towards low-scale inflation can come from the constraints on isocurvature perturbations sourced by (QCD) axionic dark matter (see e.g.~\cite{Planck2018Inflation,Axion_Cosmology_Review_Marsh:2015xka, ALPs_isocurvature_Diez-Tejedor:2017ivd}). If the Peccei-Quinn symmetry is broken during inflation, axions source dark matter isocurvature perturbations which are stronger for higher $H$ (for any given axion decay constant, $f_a$), the non-observation of which thus prefers low-scale inflation.
Furthermore, with current and future collider experiments, such as a future
$\sim \mathcal{O}(100)$ TeV collider, we might have the opportunity to investigate the physics during and after such a low-scale inflation in laboratory searches too, along with the cosmological ones!

\appendix

\acknowledgments

We are grateful to Anson Hook for useful conversation.
KD and RS are supported in part by the NSF grant PHY-1914731 and by the Maryland Center for Fundamental Physics. SK is supported in part by the NSF grants PHY-1914731, PHY-1915314 and the U.S. DOE Contract DE-AC02-05CH11231.


\bibliographystyle{JHEP}
\bibliography{Twinflation}

\end{document}